%% file: draft_coop_CIPT_1015.tex
\documentclass[11pt, journal,onecolumn,draftclsnofoot]{IEEEtran}
\usepackage{float}
\usepackage{amsmath}
\usepackage{snapshot}

\ifCLASSINFOpdf
   \usepackage[pdftex]{graphicx}
\else
   \usepackage[pdftex]{graphicx}
\fi
\begin{document}
%
\title{Wireless Video Multicast with Cooperative and Incremental Transmission of Parity Packets}
%
%
%

      
\ifCLASSOPTIONtwocolumn
\author{Zhili~Guo,~\IEEEmembership{Student Member,~IEEE,}
 Yao~Wang,~\IEEEmembership{Fellow,~IEEE,}
  Elza~Erkip,~\IEEEmembership{Fellow,~IEEE,}
   Shivendra~Panwar,~\IEEEmembership{Fellow,~IEEE,}}
\else
\author{Zhili Guo, Yao Wang, Elza Erkip, Shivendra Panwar}
   \fi

\maketitle

\begin{abstract}
\boldmath
In this paper, a cooperative multicast scheme that uses Randomized Distributed Space Time Codes (R-DSTC), along with packet level Forward Error Correction (FEC), is studied. Instead of sending source packets and/or parity packets through two hops using R-DSTC as proposed in our prior work, the new scheme delivers both source packets and parity packets using only one hop. After the source station (access point, AP) first sends all the source packets, the AP as well as all nodes that have received all  source packets together send the parity packets using R-DSTC. As more parity packets are transmitted, more nodes can recover all source packets and join the parity packet transmission. The process continues until all nodes acknowledge the receipt of enough packets for recovering the source packets. For each given node distribution, the optimum transmission rates for source and parity packets are determined such that the video rate that can be sustained at all nodes is maximized. This new scheme can support significantly higher video rates, and correspondingly higher PSNR of decoded video, than the prior approaches. Three suboptimal approaches, which do not require full information about user distribution or the feedback, and hence are more feasible in practice are also presented. The proposed suboptimal scheme with only the node count information and without feedback still outperforms our prior approach that  assumes full channel information and no feedback.
\end{abstract}


\begin{IEEEkeywords}
User cooperation, video multicast, incremental parity transmission, randomized distributed space time coding.
\end{IEEEkeywords}

%
\IEEEpeerreviewmaketitle

%
%
%
%


 

\input{cipt_intro}

 \ifCLASSOPTIONonecolumn
\fi
\input{cipt_review}
 \ifCLASSOPTIONonecolumn
\fi

\input{cipt_main}

 \ifCLASSOPTIONonecolumn
\fi
\input{cipt_simu}

 \ifCLASSOPTIONonecolumn
\fi
\input{cipt_con}
\ifCLASSOPTIONcaptionsoff
  \newpage
\fi




\bibliographystyle{IEEEtran}
\bibliography{IEEEabrv,./bib/bib_ref}

\ifCLASSOPTIONtwocolumn
\vspace{-10 mm}
\begin{IEEEbiography}[{\includegraphics[width=1in,height=1.25in,clip,keepaspectratio]{zguo_hp_bw}}]{Zhili Guo} (S'08) received the B.S. degree in communication engineering from Beijing University of Posts and Telecommunications, Beijing, China in 2008 and the M.S. degree in electrical engineering from Polytechnic Institute of New York University, Brooklyn in 2010. He is currently pursuing the Ph.D. degree at Polytechnic Institute of New York University, Brooklyn.

His research interests lie in the areas of multimedia signal processing and communication with special emphasis on video compression and wireless multimedia transmission.
\end{IEEEbiography}

\begin{IEEEbiography}[{\includegraphics[width=1in,height=1.25in,clip,keepaspectratio]{yao_bw}}]{Yao Wang} (M'90-SM'98-F'04) received the B.S. and M.S. degrees in electronic engineering from Tsinghua
University, Beijing, China, in 1983 and 1985, respectively, and the Ph.D. degree in electrical and computer engineering from University of California at Santa Barbara in 1990. 

Since 1990, she has been with the Electrical and Computer Engineering faculty of Polytechnic Institute of New York University, Brooklyn. Her research interests include video coding and networked video applications, medical imaging, and pattern recognition. She is the leading author of the textbook Video Processing and Communications (Englewood Cliffs, NJ: Prentice-Hall, 2001).

Dr. Wang is a co-winner of the IEEE Communications Society Leonard G. Abraham Prize Paper Award in the Field of Communications Systems in 2004. Shewas elected Fellowof the IEEE in 2004 for contributions to video processing and communications.
\end{IEEEbiography}

\begin{IEEEbiography}[{\includegraphics[width=1in,height=1.25in,clip,keepaspectratio]{elza1_bw}}]{Elza Erkip} (S'93-M'96-SM'05-F'11) received the B.S. degree in electrical and electronics engineering from the Middle East Technical University, Ankara, Turkey, and the M.S. and Ph.D. degrees in electrical engineering from Stanford University, Stanford, CA. 

Currently, she is a Professor of electrical and computer engineering at the Polytechnic Institute of New York University, Brooklyn. In the past, she has held positions at Rice University, Houston, TX, and at Princeton University, Princeton, NJ. Her research interests are in information theory, communication theory, and wireless communications.

Dr. Erkip received the National Science Foundation CAREER Award in 2001, the IEEE Communications Society Rice Paper Prize in 2004, and the ICC Communication Theory Symposium Best Paper Award in 2007. She co-authored a paper that received the ISIT Student Paper Award in 2007. She was a Finalist for The New York Academy of Sciences Blavatnik Awards for Young Scientists in 2010. Currently, she is an Associate Editor for the IEEE TRANSACTIONS ON INFORMATION THEORY. She was an Associate Editor for the IEEE TRANSACTIONS ON COMMUNICATIONS from 2006Ð2009, a Publications Editor for the IEEE TRANSACTIONS ON INFORMATION THEORY from 2006Ð2009, and a Guest Editor of the IEEE SIGNAL PROCESSING MAGAZINE in 2007. She was the technical program co-chair of WiOpt 2011, the co-chair of the GLOBECOM Communication Theory Symposium in 2009, the publications
chair of ITW Taormina in 2009, the MIMO Communications and Signal Processing Technical Area chair of the Asilomar Conference on Signals, Systems, and Computers in 2007, and the technical program co-chair of the Communication Theory Workshop in 2006.

\end{IEEEbiography}
\vfill
\begin{IEEEbiography}[{\includegraphics[width=1in,height=1.25in,clip,keepaspectratio]{shiv_1}}]{Shivendra S. Panwar} (S'82-M'85-SM'00-F'11) received the B.Tech. degree in electrical engineering from the Indian Institute of Technology, Kanpur, in 1981 and the M.S. and Ph.D. degrees in electrical and computer engineering from the University of Massachusetts, Amherst, in 1983 and 1986, respectively. 

He is a Professor in the Electrical and Computer Engineering Department at Polytechnic Institute of New York University, Brooklyn. He joined the Department of Electrical Engineering at the Polytechnic Institute of New York University in 1985. He is currently the Director of both the New York State Center for Advanced Technology in Telecommunications (CATT) and the Wireless Internet Center for Advanced Technology (WICAT), an NSF Industry-University Cooperative Research Center. His research interests
include the performance analysis and design of networks. Current work includes cooperative wireless networks, switch performance, and multimedia transport over networks. He has co-authored TCP/IP Essentials: A Lab based Approach (Cambridge, U.K.: Cambridge Univ. Press, 2004).

Dr. Panwar has served as the Secretary of the Technical Affairs Council of the IEEE Communications Society. He is a co-winner of the IEEE Communications Society Leonard G. Abraham Prize Paper Award in the Field of Communications Systems in 2004.
\end{IEEEbiography}
\fi






\end{document}

%% file: cipt_intro.tex
\section{Introduction}
%
%
%
%
%
%
%

Wireless video delivery of popular events is emerging as a high demand service. Advantages in bandwidth efficiency makes wireless multicast an ideal way to deliver popular live video content to many wireless nodes. However, variations in channel conditions between source and each receiver make wireless video multicast a challenging problem. Cooperative communication techniques effectively combat the variations in channel quality\cite{elza:TransComm}. One way to enable multiple nodes to cooperate simultaneously is by using distributed space-time codes (DSTC)\cite{laneman}. However, DSTC requires a predetermined and fixed number of relays, and requires tight coordination and synchronization among the relays. To relax these restrictions, Randomized DSTC (R-DSTC)\cite{mergen:TSP} lets each relay transmit a random linear combination of antenna waveforms, and enables all nodes to join in the relaying phase, without requiring strict coordination and synchronization. 


To compensate for packet loss during transmission, packet level forward error correction (FEC) \cite{maj:fec,joo:fec,naf:fec} and Automatic Repeat reQuest (ARQ) \cite{dia:arq} are widely used. In our previous work \cite{OZGU:tmm11} and \cite{ozgu:PIMRC10}, packet level FEC using Reed-Solomon codes was employed in conjunction with cooperative transmission.

Randomized cooperation for video multicast in an IEEE 802.11g based WLAN scheme is studied in \cite{OZGU:tmm11}, where the source (access point, or AP) transmits a video packet, and then all nodes receiving the packet forward this packet simultaneously using R-DSTC. To combat packet losses, the source sends both the original video packets (called source packets) as well as parity packets needed by the receivers for recovering lost source packets. Each packet goes through two hops. The transmission rates of both hops and the FEC rate are chosen to maximize the video rates that a large percentage of nodes can receive without errors. Throughout this paper, this scheme will be referred as \textit{multicast-RDSTC}. 

An improved parity packet transmission scheme for \textit{multicast-RDSTC} is proposed in \cite{ozgu:PIMRC10}, named \textit{enhanced-multicast-RDSTC}. In this scheme, the AP with the help of relays first transmits all the source packets using two hops. Upon the completion of \textit{k} source packet transmissions (each source packet going through two hops), the nodes that can recover all \textit{k} source packets correctly generate parity packets and transmit them using R-DSTC. As more parity packets are transmitted, more relays join in parity packet generation and transmission. Transmission rates at both hops for source packets, and the transmission rate for parity packets, are chosen to maximize the achievable video rate at all nodes. Simulations show that \textit{enhanced-multicast-RDSTC} can yield a significant increase in the video rate compared to \textit{multicast-RDSTC}, by reducing the number of hops for parity packets.

In this paper, we propose an innovative way to implement source and parity packet transmissions to further develop the potential of video multicast with R-DSTC. The AP will first transmit all source packets without using relays. After the source finishes the transmission of \textit{k} source packets (each using only one hop), the source will start to generate and transmit parity packets. Nodes which receive all \textit{k} source packets will also join in the generation and transmission of the first parity packet. More nodes will join in the generation and transmission of additional parity packets as soon as they receive a total of $k$ packets (source or parity), and can therefore decode all $k$ source packets. The parity transmission will stop only after all nodes receive at least $k$ packets. To inform all users in the system whether additional parity packets are needed, we further propose an efficient feedback mechanism. After transmission of all source packets, and after each additional parity  packet, nodes which still have not received at least $k$ packets will simultaneously send a request for more parity packets using R-DSTC. For each user distribution, we determine the optimal transmission rate for each phase to maximize the achievable video rate. We refer to this new scheme  as \textit{CIPT-multicast-RDSTC}, where CIPT stands for cooperative incremental parity transmission. Schemes that require less information about user distribution or feedback are also considered, which are suitable for practical deployment.


Simulation results show that the \textit{CIPT-multicast-RDSTC} scheme assuming full channel information and feedback provides significant gains over the previous \textit{multicast-RDSTC} scheme, increasing the achievable rate by 27\% on average over a range of node counts from 16 to 80. Compared to the \textit{enhanced-multicast-RDSTC}, which also requires feedback, it increases the rate by 15\% on average. Even with only node count information and without feedback, the new scheme provides about 17\% increase in rate over the previous \textit{multicast-RDSCT} scheme that requires full channel information but no feedback, and provides a gain of 30\% over the multicast-RDSTC scheme that requires only the node count information.

%

This paper is organized as follows. In Section \ref{sec:review}, our previous work on cooperative video multicast scheme is reviewed. We introduce our proposed system in Section \ref{sec:cipt}. The simulation setup and results are presented in Section \ref{sec:simulation}. We conclude the paper in Section \ref{sec:con}.

%

%% file: cipt_review.tex
\section{Review of Previous Video Multicast Scheme with R-DSTC}
\label{sec:review}

\begin{table*}
\caption{Notation}\label{tab:notation}
\begin{center}
\resizebox{\textwidth}{!}{
\begin{tabular}{ |l|l||l|l| }
  \hline
  $R_1$ & First hop transmission rate for RDSTC multicast (bits/sec) &   $L$ & Space time code dimension \\
  & First hop transmission rate for the source packets for enhanced-RDSTC-multicast &  $\gamma$ & FEC rate\\
  $R_2$ & Second hop transmission rate for RDSTC multicast (bits/sec) &$k$ & Total number of source packets per FEC block\\
 &  Second hop transmission rate for the source packets for enhanced-RDSTC-multicast & $m$ & Parity packets needed per FEC block\\
  $R_s$ & Source packets transmission rate for CIPT multicast (bits/sec) & $N_T$ & Total number of nodes \\
  $R_p$ & Parity packets transmission rate for CIPT multicast (bits/sec) & $R_v$ & Achievable video rate (bits/sec) \\
 & Parity packets transmission rate for enhanced-RDSTC-multicast & & \\
    \hline
\end{tabular}}
\end{center}
\end{table*}

\begin{figure}[!htbp]
\centering
\ifCLASSOPTIONtwocolumn
\includegraphics[width=0.95\columnwidth]{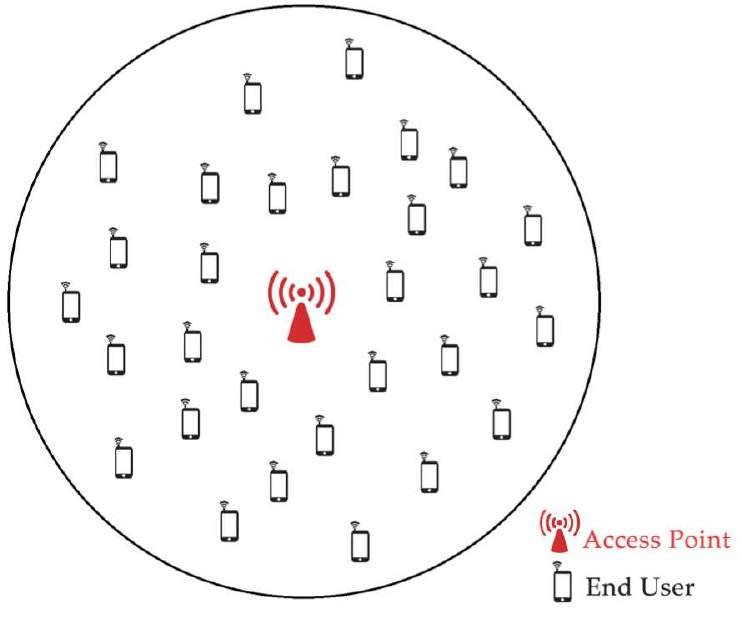}
\else
\includegraphics[width=0.55\columnwidth]{illu_sys_start_1}
 \fi
\caption{Multicast system layout. The AP is located at the center, with end users randomly placed around the AP within its coverage area, which is a circle of radius $r$.}
\label{fig_sys_start}
\end{figure}

This section summarizes our previous work \cite{OZGU:tmm11}\cite{ozgu:PIMRC10}, where we studied video multicast within an infrastructure-based wireless network. The AP transmits the video to all nodes in a multicast session within its coverage range as shown in Figure \ref{fig_sys_start}. We assume that all channels between the receiving nodes and the AP and between the nodes undergo independent slow Rayleigh fading, and the fading level is constant over the duration of single packet transmission. We further assume that a pre-determined set of transmission rates can be achieved using different modulation and channel coding schemes. We refer the readers to \cite{OZGU:tmm11} for details.

\subsection{Multicast-RDSTC}
In the Multicast-RDSTC scheme, the AP generates $m$ parity packets for every block of $k$ source packets and transmits each of these packets at a transmission rate of $R_1$ bits/sec. Nodes that receive each packet (either a source packet or a parity packet) correctly, relay this packet simultaneously using R-DSTC with STC dimension $L$ (which is the number of antennas used for the STC) to other nodes at a transmission rate of $R_2$ bits/sec. In order not to increase the total radiated power over the air, each relay transmits with a power that is equal to the transmission power of the AP divided by the average number of relays for a given node count (the number of users in the multicast session). We assume such information can be predetermined through simulation \cite{OZGU:tmm11}.

Note that there are only a few options for the STC dimension $L$ due to the limited dimensions of practical STC codes. Each $L$ has a corresponding STC code rate which will affect effective transmission rate. A full rate can be realized only when $L = 2$ \cite{OZGU:tmm11}. Therefore, in all results reported here, we use $L=2$. Optimization of the other parameters is discussed in Sec. \ref{rdstc_nc}.


\subsection{Enhanced approach for multicast-RDSTC}

\begin{figure}[htbp]
 \begin{center}
 \ifCLASSOPTIONtwocolumn
  \includegraphics[width=0.75\columnwidth]{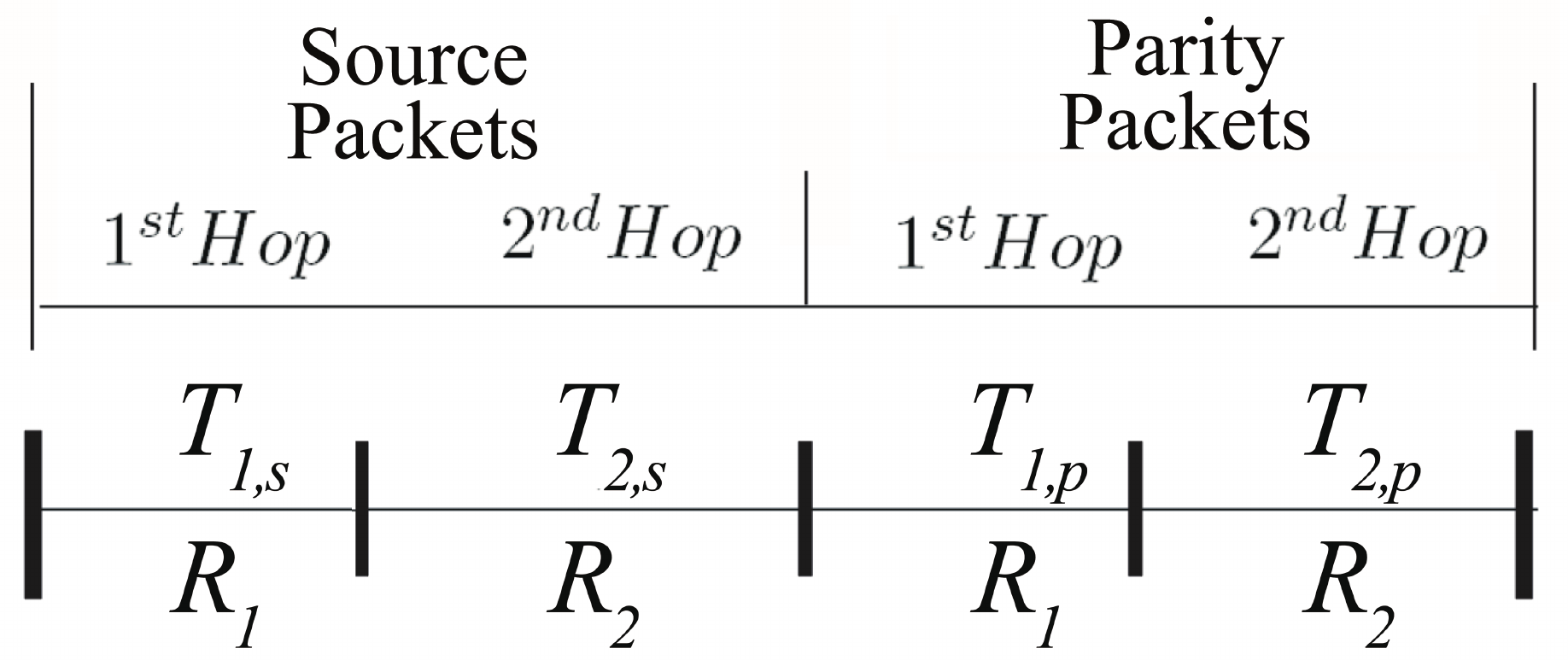} 
  \\(a) RDSTC scheme \cite{OZGU:tmm11}
  \\
  \vspace{10pt}
  \includegraphics[width=0.75\columnwidth]{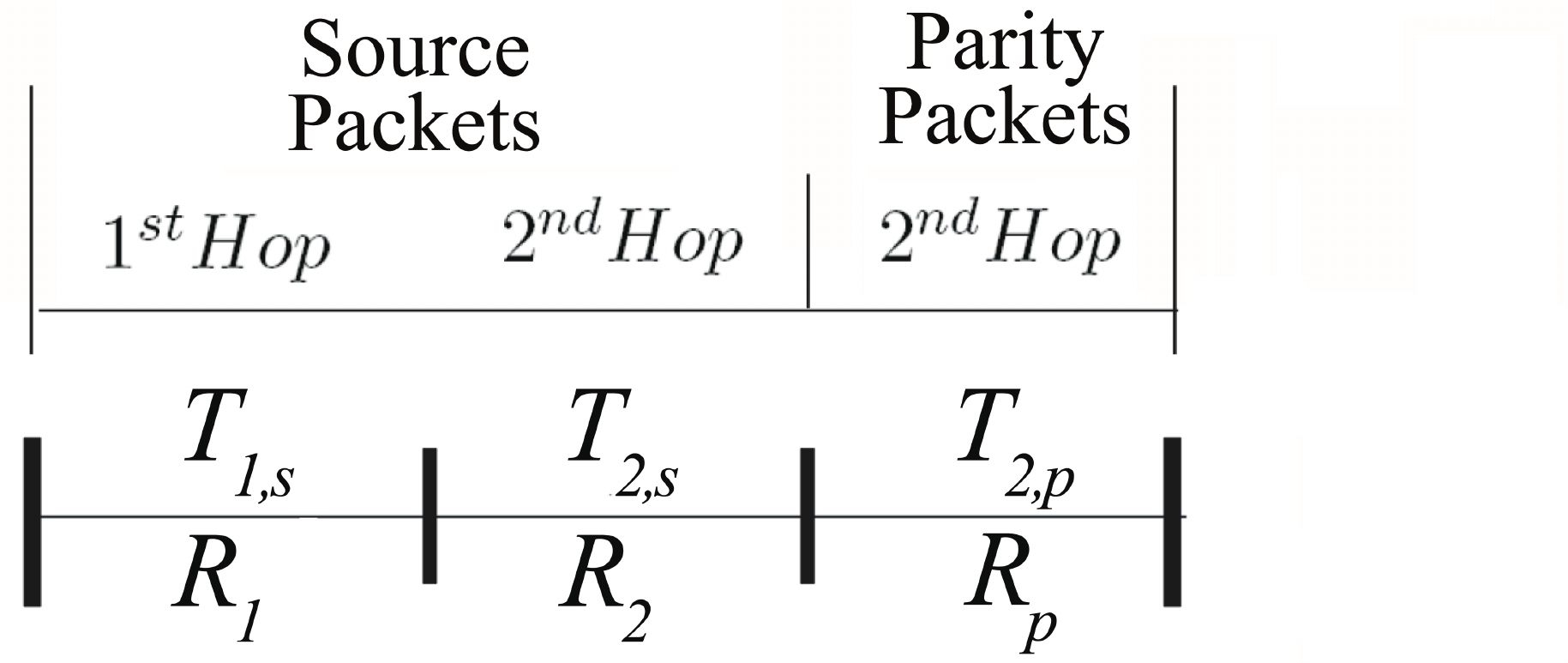} 
  \\(b) Enhanced RDSTC scheme \cite{ozgu:PIMRC10}
   \\
  \vspace{10pt}
  \includegraphics[width=0.75\columnwidth]{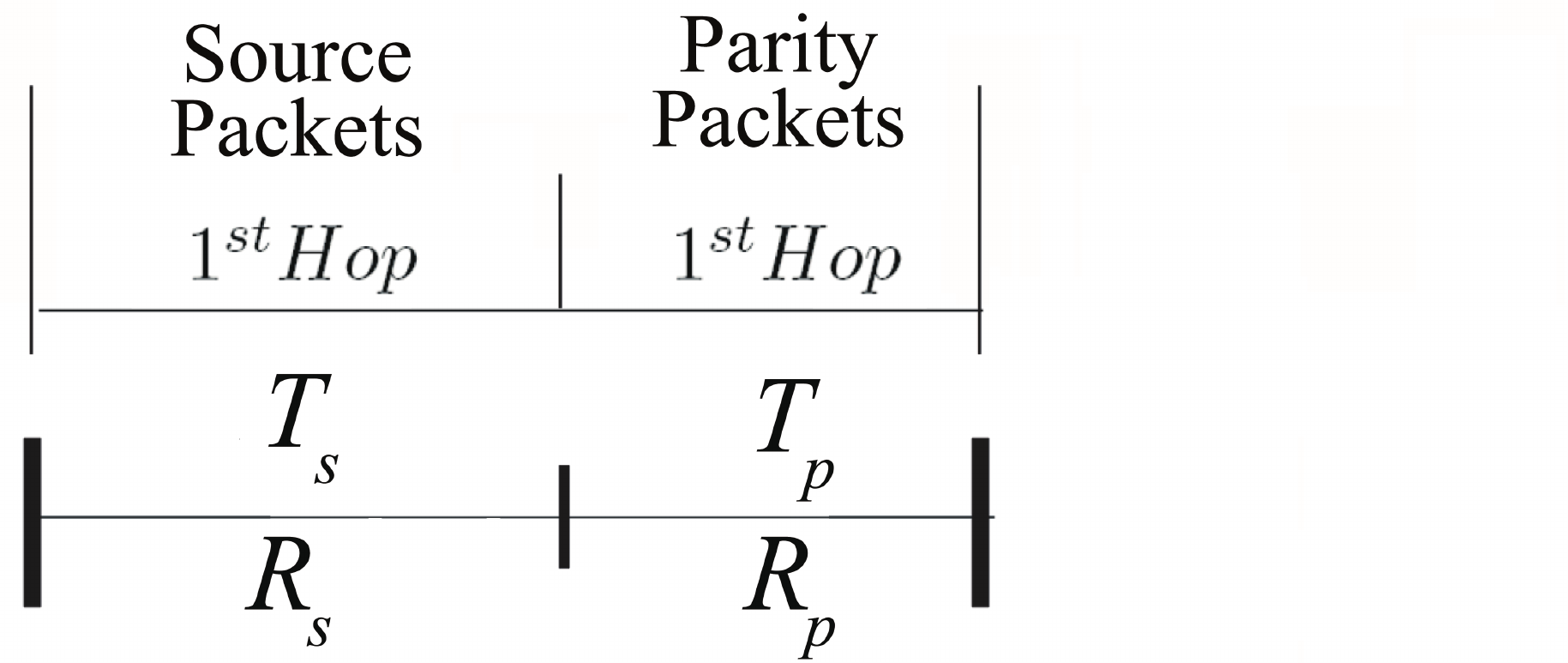} 
  \\(c) CIPT RDSTC scheme proposed in this paper
\else
  \includegraphics[width=0.4\columnwidth]{fig_4_1_new} 
  \\(a) RDSTC scheme \cite{OZGU:tmm11}
  \\
    \vspace{10pt}
  \includegraphics[width=0.4\columnwidth]{fig_4_2_new} 
  \\(b) Enhanced RDSTC scheme \cite{ozgu:PIMRC10}
  \\
    \vspace{10pt}
  \includegraphics[width=0.4\columnwidth]{fig_4_3_new} 
  \\(c) CIPT RDSTC scheme proposed in this paper

 \fi

 \caption{Transmission rates and time scheduling for different multicast schemes using RDSTC}\label{fig:review_1and2}
 \end{center}  
\end{figure}

Multicast-RDSTC and enhanced-multicast-RDSTC \cite{ozgu:PIMRC10} differ in the way parity packets are sent. For multicast-RDSTC, the AP transmits both the source and parity packets at rate $R_1$, and each packet is then relayed once at rate $R_2$ by nodes that receive the packet. For the enhanced-multicast-RDSTC scheme, the AP is only responsible for the transmission of the source packets. The AP transmits the source packets at a transmission rate of $R_1$ bits/sec and the relays forward these packets using R-DSTC at a rate $R_2$ bits/sec. After the completion of $k$ source packet transmission, the nodes that receive all $k$ source packets become parity relays. The parity relays generate parity packets and transmit them using R-DSTC at a rate $R_p$ bits/sec. Note that after each parity packet transmission, any node that receives a total of $k$ packets out of all packets transmitted so far, can decode to obtain the $k$ source packets, becomes a parity relay, and joins the parity packet transmission. Therefore, the number of parity relays increases with time. Fig. \ref{fig:review_1and2} illustrates the time scheduling of \textit{multicast-RDSTC} and \textit{enhanced-RDSTC}.

\subsection{Optimization Based on Channel Information}
\label{rdstc_nc}
In the ``full channel'' case of Multicast-RDSTC, the system requires message exchange between the AP and nodes, and between pairs of nodes, to determine the channel quality between the AP and each node, and between all nodes, which can be described by a channel quality matrix. For each particular node placement (with a corresponding channel quality matrix), we examine different candidates of transmission rate combinations ($R_1$,$R_2$). For each candidate ($R_1$,$R_2$), we can determine the maximum end-to-end PER (averaged over fading) among all nodes. We then determine the suitable FEC rate to ensure that the maximum FEC decoding failure probability among all nodes is less than $\tau$. Our own perceptual observation is that as long as the FEC decoding failure probability is below threshold $\tau=0.5\%$, the loss effect is hardly noticeable, the decoded video quality is almost equal to the encoded video quality which in turn depends only on the video rate\cite{OZGU:tmm11}. This observation can also be validated by the model in \cite{REI:pl10}.

Given the transmission rates and its corresponding FEC rate, we can determine the achievable video rate. The transmission rate candidate and its corresponding FEC rate that yields the largest video rate constitutes the optimal operating point for this channel quality matrix. We assume that such optimal operating points can be predetermined by simulations for various likely full channel quality matrices, and saved in the AP.  The AP periodically collect channel quality information to update its channel quality matrix, and consequently the operating point. 

We also considered a simplified realization for Multicast-RDSTC, referred as ``multicast-RDSTC-node-count''. It does not require the full channel information, but only the node count of the multicast session. To determine the optimal operating point for a given node count, we randomly generate multiple node placements where nodes are uniformly distributed. For each candidate transmission rate combination ($R_1$,$R_2$), we do not consider the worst 5\% node placements with worst maximum Packet Error Rates (PER). We find the maximum average PER among all remaining 95\% node placements and compute the corresponding $\gamma$ and hence the video rate based on this PER. We choose ($R_1$,$R_2$) and the corresponding $\gamma$ to maximize the video rate. In practice, a table of the system operating parameters ( $R_1$, $R_2$, and $\gamma$) for different node counts can be pre-computed and stored at the source station.

For the \textit{enhanced-RDSTC} system, we only considered the version assuming full-channel information. This system needs users to send feedback to help the AP decide when to start transmitting new FEC blocks. This type of feedback information is also needed in the proposed CIPT system, and  will be discussed in Section \ref{sec_main_fb}.
%


%% file: cipt_main.tex
\section{Proposed Video Multicast Scheme with Incremental Parity Packets}
\label{sec:cipt}
\subsection{System Model Overview}

In contrast to \textit{multicast-RDSTC} and \textit{enhanced-multicast-RDSTC}, for the \textit{CIPT-multicast-RDSTC} scheme, the source packets will be sent only once from the AP at transmission rate $R_s$ bits/sec and will not be relayed. Let us denote the source packets transmission as \textit{phase 1}. After \textit{phase 1} is completed, the parity transmission starts, which is \textit{phase 2}. 

\begin{figure*}[!htbp]
\centering
\includegraphics[width=0.95\textwidth]{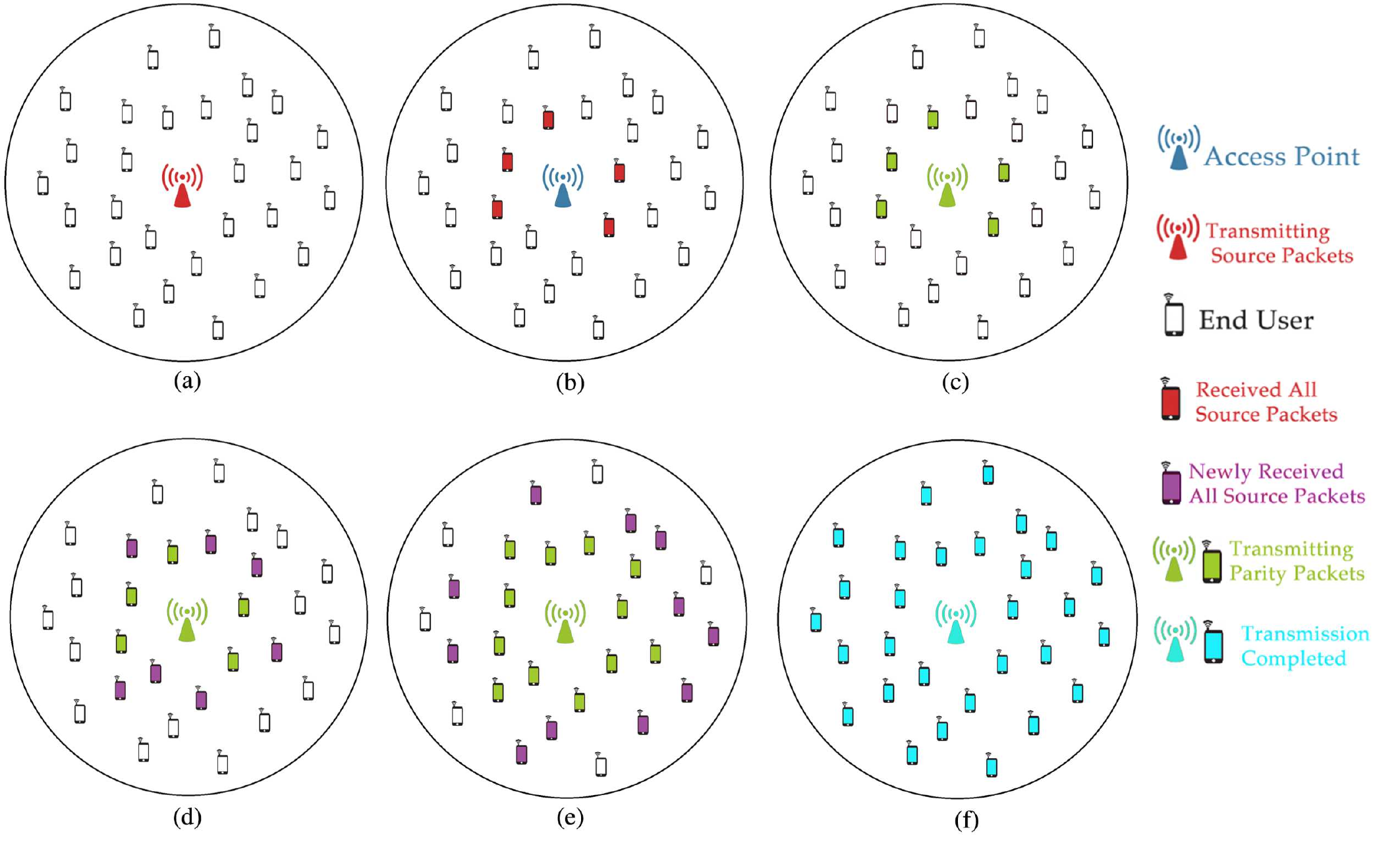}
\caption{User status during one FEC block (with $k$ source packets) transmission for CIPT system: (a) The AP transmits all $k$ source packets; (b) after source packets transmission, several users received all $k$  source packets; (c) users who received all $k$ source packets generate and transmit first parity packet along with AP simultaneously using RDSTC (incremental parity packets transmission started); (d) Users who receive a total of $k$ packets will recover the original  $k$ source packets, and join the previous relay set and the source to generate and simultaneously transmit the second parity packet; (e) as more parity packets are transmitted, more users will be able to correctly recover all source packets and join the parity transmission; (f) once every user received $k$ packets, this FEC block is completed.}
\label{fig_relay}
\end{figure*}

The parity packets are generated and transmitted incrementally.  The process is illustrated in Fig. \ref{fig_relay}. In the beginning of \textit{phase 2}, only nodes which receive all source packets join the AP to generate and send the first parity packet. Nodes who receive a total of $k$ packets (out of $k$ source packets, and the first parity packet) can now recover $k$ source packets and join the previous parity transmission group to generate and send the second parity packet. As more parity packets are transmitted, more nodes join the parity transmission group. This process  continues until all nodes receive $k$ packets out of all transmitted source and parity packets. Note that we assume there is a short feedback phase after transmission of each parity packets, for users who have not received  $k$ packets to indicate that they need to receive more packets. All parity packets will be sent using R-DSTC at transmission rate $R_p$ bits/sec. Both the AP and the other parity relays transmit at a power that is equal to the AP transmission power for the source packet divided by the expected number of the parity transmission nodes (to be estimated from prior simulation studies). Note that if there is no available \textit{parity relay} at the beginning of \textit{phase 2} (which will often be the case for systems with low user density), the AP will multicast parity packets by itself. With the previous \textit{enhanced-multicast-RDSTC} scheme, because the AP does not join parity packet transmission, we had to use two-hop transmission for the source packets  to ensure a sufficient number of parity transmission nodes after the completion of source packet transmission.

For the proposed \textit{CIPT-multicast-RDSTC} scheme, the number of parity packets $m$, and hence the FEC rate, depends on both the source transmission rate $R_s$ and the parity transmission rate $R_p$. Therefore, $m$ and hence the FEC rate $\gamma$, is a function of $R_s$ and $R_p$.

Assume an average packet size of $B$. Then the transmission time for sending $k$ source packets at transmission rate $R_s$ is $T_s = kB/R_s$. Similarly, the transmission time for sending $m$ parity packets at transmission rate $R_p$ is $T_p = mB/R_p$. The video rate $R_v$ is therefore:
\begin{equation}
R_v = \frac{\beta kB}{T_s+T_p}=\frac{\beta kR_sR_p}{kR_p+mR_s}=\frac{\beta\gamma R_sR_p}{\gamma R_p + (1-\gamma)R_s}
\label{eq:RV_CIPT}
\end{equation}
where $\beta$ denotes the effective data ratio, defined as the ratio of the data rate used to transmit video data to the total sustainable rate.

\subsection{Implementation of Channel Information Update and Feedback}
\label{sec_main_fb}
There are two types of message exchange that are needed in the CIPT-multicast-RDSTC system. The first one is needed for updating the channel information, which consists of all signaling signals needed to deduce the channel quality in terms of the channel SNR between the source and all the nodes and between all pairs of nodes. In order for the source station to know the average channel qualities among the nodes, the nodes could exchange control signals among themselves to measure the average SNR, and then transmit this information back to the source station. Because the channel information is used to determine the operating parameters for transmitting each FEC block, we envision that such an update to be done at the beginning of every new FEC block or every several FEC blocks, depending on the expected channel dynamics. 

According to the updated channel quality matrix, the channel is classified into one of several pre-determined channel states each representing a cluster of channel conditions with similar channel quality matrices. Each state is described by the average channel quality matrix of the cluster.   We  assume that through pre-simulations, the optimal operating points in terms of $R_s$, $R_p$ and possibly the required number of parity packets $m$ are determined and saved for all possible channel states. The information regarding the average number of available relays at each additional parity packet transmission time is also predetermined and saved, to enable power normalization for the parity transmission. Once the AP updates its channel state, it broadcasts the optimal operating point as well as the desired parity transmission powers at different times to all nodes, so that all nodes who can join parity transmission at some time can use the correct transmission rate and power levels. 

The second type of message exchange is needed to inform the system when to stop transmitting more parity packets. We assume that by the end of the transmission of every new parity packet, there will be a pre-allocated time slot for users who have not received $k$ packets to multicast a feedback to indicate that at least one user still needs to receive parity packets. Thanks to the nature of R-DSTC, users who have not received $k$ packets can multicast the feedback signal simultaneously using only one time slot. The AP and nodes who have received $k$ packets will listen for this feedback signal at the end of each parity packet transmission, and will keep generating and sending additional packets until no feedback signal is received. 

The feedback signal can be any short packet that can be sent over the designated feedback slot. For a typical WiFi environment, for instance, under 802.11g \cite{802_11g} framework, we could use the PLCP Preamble as our desired feedback message, which will take only $B_{fb} = 72 bits = 9 bytes$. To guarantee that the feedback packet can be received correctly by nodes everywhere in the multicast session, we use the lowest transmission rate possible, which is $R_{fb} = 6 Mbps$ in the 802.11g environment. The transmission time for $m$ feedback packets is $T_{fb}=mB_{fb}/R_{fb}$. The video rate considering feedback can be derived as:
\begin{equation}
\ifCLASSOPTIONtwocolumn
\begin{aligned}
R_{v_{fb}} & = \frac{\beta kB}{T_s+T_p+T_{fb}} \\
& =\frac{\beta kR_sR_pR_{fb}}{kR_pR_{fb}+mR_sR_{fb}+\alpha mR_sR_p}\\
& =\frac{\beta\gamma R_sR_p}{\gamma R_p + (1-\gamma)R_s + \alpha(1-\gamma)\frac{R_sR_p}{R_{fb}}}
\end{aligned}
\else
R_{v_{fb}} = \frac{\beta kB}{T_s+T_p+T_{fb}}=\frac{\beta kR_sR_pR_{fb}}{kR_pR_{fb}+mR_sR_{fb}+\alpha mR_sR_p}=\frac{\beta\gamma R_sR_p}{\gamma R_p + (1-\gamma)R_s + \alpha(1-\gamma)\frac{R_sR_p}{R_{fb}}}
\fi
\label{eq:RV_fb_CIPT}
\end{equation}
where $\alpha=B_{fb}/B$. The rate reduction factor due to feedback is thus
\begin{equation}
\omega = \frac{R_{v_{fb}}}{R_v}=\frac{(1-\gamma)\rho_s+\gamma\rho_p}{\gamma\rho_p+(1-\gamma)(1+\alpha\rho_p)\rho_s}
\label{eq:R_fb_f}
\end{equation}
where $\rho_s=R_s/R_{fb}$ and $\rho_p=R_p/R_{fb}$.



If we assume the regular video packet size is $B=1400 bytes$ and the feedback packet size $B_{fb}=9 bytes$, then $\alpha=0.0064$. For our proposed system, the optimal operating point for most user configurations consists of $R_s = 24 Mbps, R_p=36 Mbps$ and $\gamma\approx0.5$. With $R_{fb} = 6 Mbps$, we have $\rho_s = 4$ and $\rho_p = 6$. With these assumptions, we get $\omega = 0.9848$, meaning that the feedback has about 1.5\% total overhead.

%
%
%
%
%
%
%

\subsection{Optimization of Transmission Rates}
\label{cipt_nc}
By switching on and off the above two types of message exchanges, we will have four variations of the CIPT-multicast system. We will discuss optimization of the transmission rates for each variation in this subsection.

\subsubsection{Full channel information with feedback}
\label{cipt_full_fb}
This scenario assumes the AP periodically updates the full channel information to determine the channel state, as described in Sec. \ref{sec_main_fb}. Furthermore, it assumes that users are able to multicast feedback packets to make the system determine the termination time for each FEC block. 

To determine the optimal operating point for each channel state,  we go through all possible pairs of feasible $R_s$ and $R_p$. For each candidate pair of $R_s$ and $R_p$,  through channel simulation, we determine the necessary number of parity parity $m$ for the transmission of each FEC block, so that all nodes receive at least $k$ packets. We find the average of $m$ over many FEC blocks. We use this $m$, in addition to  $R_s$ and $R_p$ to determine the video rate using Eq.(\ref{eq:RV_fb_CIPT}). Finally we choose the optimal $R_s$ and $R_p$ that maximizes the video rate.




The optimal $R_s$ and $R_p$ for different channel states are stored in a lookup table. Note that $m$ does not need to be stored in the look up table, as we use feedback to determine the necessary $m$ for each particular FEC block.

\subsubsection{Full channel information with no feedback}
We assume that full channel information is periodically updated and available to the AP in this system configuration. However, feedback is disabled for easier and more practical implementation. The video rate $R_v$ is computed by using Eq.(\ref{eq:RV_CIPT}).

For each channel state, at every feasible operational rate pair ($R_s$,$R_p$), the necessary $m$ to satisfy a FEC decoding failure threshold $\zeta$ is predetermined and recorded. To calculate the FEC decoding failure rate when $m$ parity packets are transmitted, we first obtain the FEC decoding failure rate for each user averaged over many different fading realizations. Then we pick the largest FEC decoding failure rate among all users for the same $m$. Finally we choose an $m$ such that the largest decoding failure rate among all users is less than a preset threshold $\zeta$. Then, similar to the previous scenario, the source chooses the optimum $R_s$,$R_p$, and the corresponding $m$ that maximize the video rate.  As with the previous RDSTC-multicast scheme for determining the necessary parity packet number, we set $\zeta=0.5\%$.

Unlike the previous scenario, this time $m$ for different channel states will also be stored in a lookup table to let AP and all the users know when to terminate the transmission of the current FEC block without utilizing feedback. 

\subsubsection{Node-Count information with feedback}
In this case, we assume the AP only knows the node count. It still requires feedback from all nodes to determine when the parity transmission should be terminated. For each given node count,  we generate multiple node placements each with uniform distribution of nodes in the coverage area. As with the node-count version in the \textit{multicast-RDSTC} system \cite{OZGU:tmm11}, for each feasible pair of ($R_s$, $R_p$), we remove the worst 5\% of node placements. Specifically, we find the parity packet number $m$ needed for each node placement. We remove 5\% of node placements with largest $m$. Then we find the maximum parity packet number $m^*$ needed among all remaining node placements. We compute the video rate corresponding to this $m^*$ and the rate pair ($R_s$, $R_p$).  Finally we choose the rate pair ($R_s^*$, $R_p^*$) that achieves the highest video rate as the optimal operating point for this node count.

In practice, a table of the system operating parameters ($R_s^*$, $R_p^*$), for different node counts can be pre-computed and stored at the AP. Note that in the system operation, the necessary $m$ for a particular node placement for each FEC block is determined based on the actual feedback, and is in most cases smaller than the one  used for determining the optimal transmission rates. 

\subsubsection{Node-Count information with no feedback}
In this  system, we assume there is no feedback to indicate whether a node has received enough packets. As with the previous case, we determine the optimal  ($R_s^*$, $R_p^*$) for each node count. But we also record the corresponding maximum number of parity packets $m^*$ for the chosen ($R_s^*$, $R_p^*$). All this information will be precomputed and stored in a look-up table. As is apparent, this system will be the easiest one to implement.

%% file: cipt_simu.tex
\section{Simulation Results}
\label{sec:simulation}
We have simulated video multicast using different versions of the CIPT transmission scheme, as well as direct transmission, RDSTC-multicast\cite{OZGU:tmm11} and enhanced-RSTC-multicast\cite{ozgu:PIMRC10} for comparison. In our simulations, we generate 300 random node placements for each node count, and report the average performance over these 300 node placements for each node count. There are five node counts considered, which are 16, 32, 48, 64, and 80 nodes. We assume that each transmission block has $k$ source packets. In most of our results, the FEC block size is set at $k=64$. We have also examined other possible values of $k$, including 4, 8, 16, and 32, to evaluate the impact of $k$ on the system performance. 

For each particular node placement for a given node count, we simulate the transmission of 20,000 blocks, with independent fading realizations among these blocks, and report the average performance over these blocks. We keep the fading level constant within each packet and independent among all packets. Through channel simulation, we determine whether a source packet send by the AP only using a given source transmission rate is received  by each node, and whether a parity  packet sent by the AP and relay nodes using RDSTC at an assumed parity transmission rate is received by each remaining node.  The simulation procedure can be found in \cite{OZGU:tmm11}. 

For the systems assuming full channel information (including CIPT and RDSTC-multicast), to find the optimal operating point (in terms of transmission rates for different transmission phases and corresponding FEC rate) for a particular node placement, we treat each node placement as a separate channel state, and use the method described in Sec.\ref{cipt_full_fb} to determine the optimal operating point for each state. For the  systems that assume only the knowledge of the node count (including CIPT-nodecount, CIPT-nodecount-nofeedback, and RDSTC-nodecount),  for a given node count, among the 300 node placements, we identify 5\% of those (i.e., 15 node placements) that have the worst end-to-end packet error rate (PER) (for systems derived from CIPT, it means largest $m$), and determine the optimal operating point for the remaining 95\% node placements as described in Section \ref{rdstc_nc} and \ref{cipt_nc}. 


\begin{figure}[!h]
\centering
\ifCLASSOPTIONtwocolumn
\includegraphics[width=0.95\columnwidth]{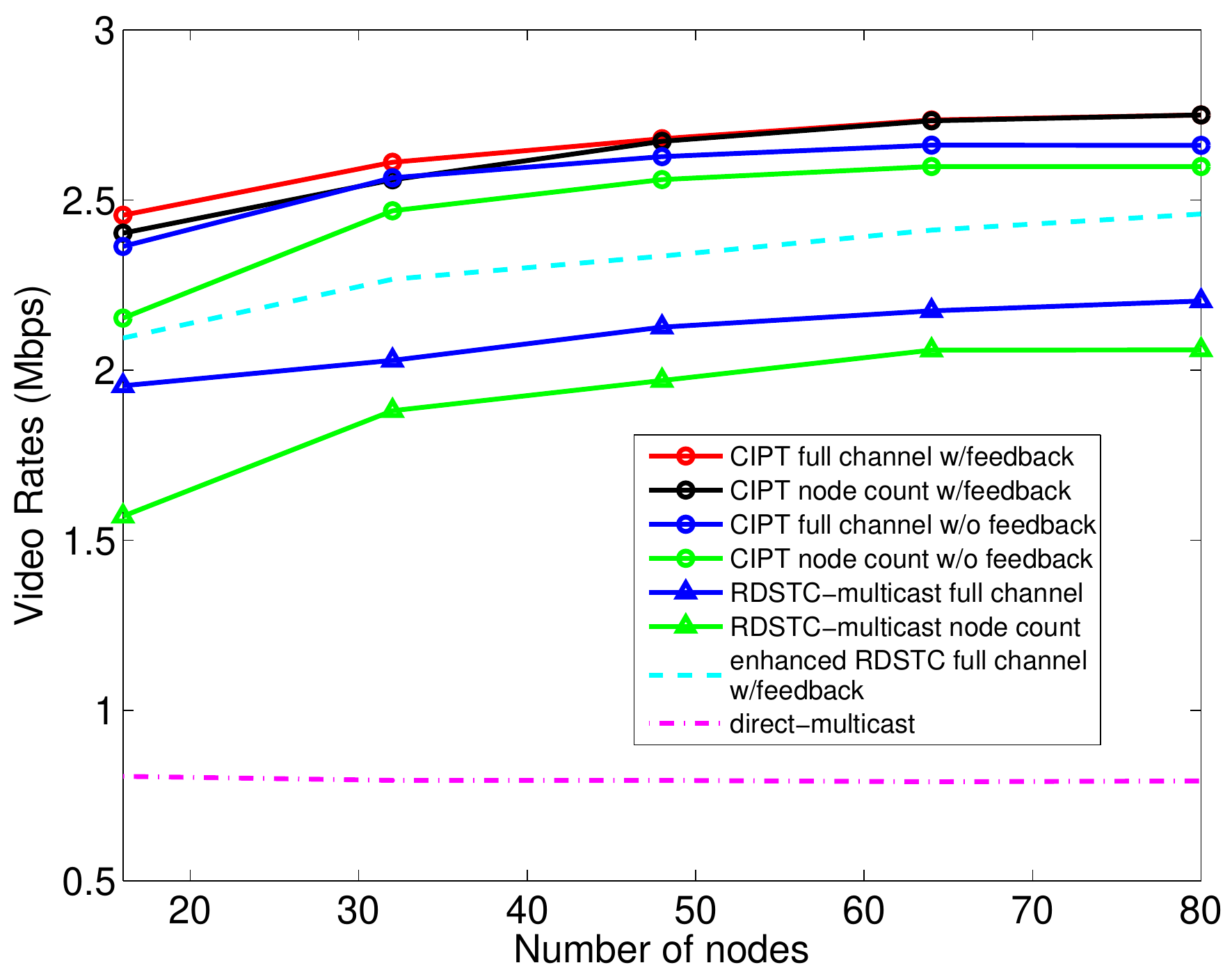}
\else
\includegraphics[width=0.65\columnwidth]{rates_NEW_1116}
 \fi
\caption{Achievable video rates of different cooperative video multicast schemes. Note that curves with the same color indicate systems with the same requirement regarding feedback and channel information. On the other hand, the same marker indicates that those curves are using the same transmission scheme. Number of source packets per FEC block $k=64$.}
\label{fig_sim}
\end{figure}

Figure \ref{fig_sim} compares the achievable video rates by different transmission schemes. For all schemes,  the achievable video rate increases with the node count. This is because, with more users in a multicast session, thanks to the nature of cooperative communication, we can have more nodes available to help in the parity packets transmission phase, which reduces the number of parity packets required.  

Figure \ref{fig_sim} shows that CIPT scheme outperformed multicast-RDSTC and enhanced-RDSTC-multicast in every comparable situation. For example, CIPT full channel w/o feedback can increase the video rate by 22\% over the RDSTC-multicast full channel, both requiring full channel information, but no feedback. When both are using only node count information, CIPT node count w/o feedback improves over RDSTC node count by 30\%. Furthermore,  CIPT full channel w/feedback outperforms enhanced RDSTC full channel w/feedback by 15\%. 

Among the CIPT systems, compared to CIPT with full channel information and feedback (denoted by CIPT full channel w/feedback), using node count information only but retaining feedback (denoted by CIPT node count w/feedback) leads to small degradation when the node count is small. As the node count increases, the difference becomes negligible. This is as expected; with more nodes, which are uniformly distributed, the optimal operating point depends less on the actual node placement. When the system still assumes full channel information, but does not require feedback (CIPT full channel w/o feedback), the performance degradation is larger, and remains non-negligible even for larger node counts. This is because feedback enables the system to provide just enough parity packets for each FEC block. Without feedback, the system has to choose $m$ in a conservative fashion to guarantee that all users receive at least $k$ packets within each transmission block. However, as demonstrated in Fig.\ref{fig_sim}, the degradation due to the removal  of feedback is relatively small, with a video rate reduction of 4\%, slightly more with fewer nodes. Comparison of ``CIPT full channel w/o feedback'' and ``CIPT node count w/feedback'' reveals that the feedback information is more important than the full channel information. When we  remove the requirement for both the full channel information and  feedback (i.e. ``CIPT node count w/o feedback''), the system performance is decreased further, with a video rate decrease of 5\%-12\% compared to the full-channel with feedback case. However, such performance degradation may be well justified in practice, because this simplified system does not require any feedback or message exchange among the users and the source station.  Even with this most simplified CIPT system, the video rate increases by 17\% over  the previous RDSTC-multicast system using  full channel information.

\begin{figure}[!h]
\centering
\ifCLASSOPTIONtwocolumn
\includegraphics[width=0.95\columnwidth]{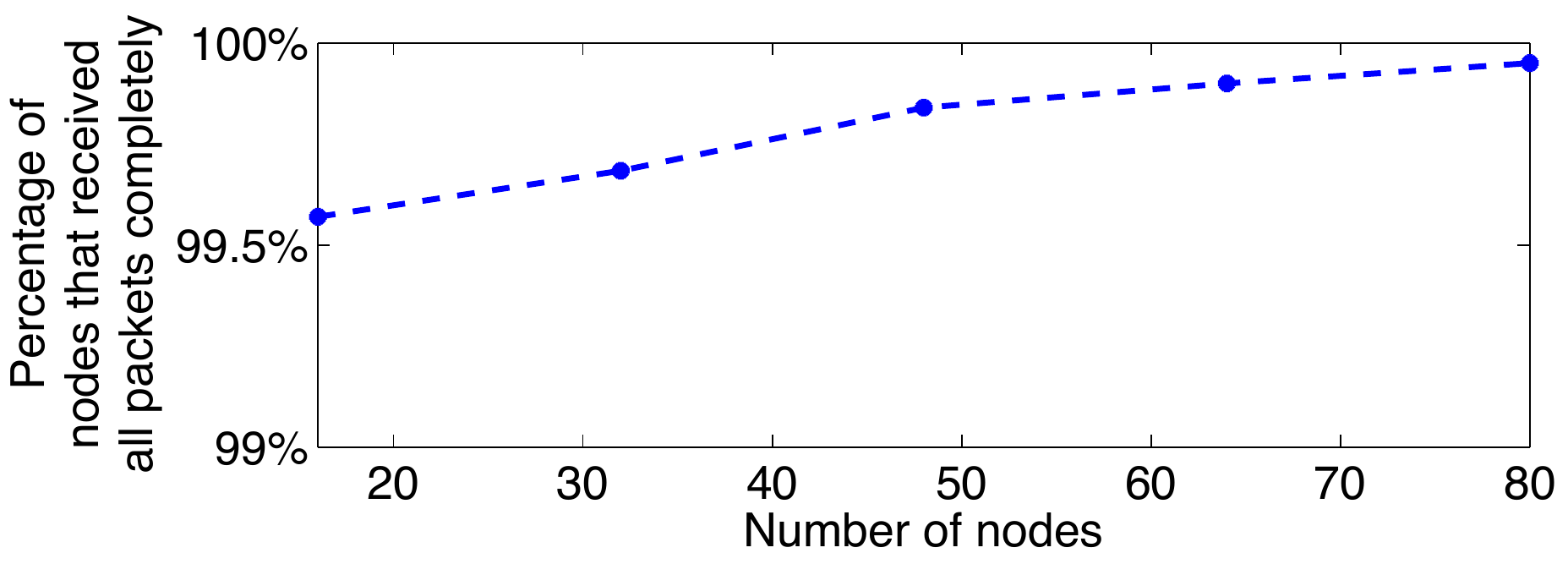}
\else
\includegraphics[width=0.65\columnwidth]{percentage_movid_1}
 \fi
\caption{Percentage of nodes that receive all packets at all node placements versus number of nodes in the CIPT node count w/o feedback.}
\label{fig:per_receive}
\end{figure}

Recall that when using node count information to substitute the full channel information, we ignore node placements with the worst 5\% performance when determining the optimal operating point for each node count. Similarly, when foregoing the feedback information,  we choose  the number of parity packets to be sent so that each user will see at most 0.5\% FEC decoding failure rate. Figure \ref{fig:per_receive} shows that for the CIPT node count w/o feedback system, among all nodes in all node placements considered (including worst 5\%), the percentage of users that can not recover all source packets is  only less than 0.43\%. 

\begin{figure}[h]
\centering
\ifCLASSOPTIONtwocolumn
\includegraphics[width=0.95\columnwidth]{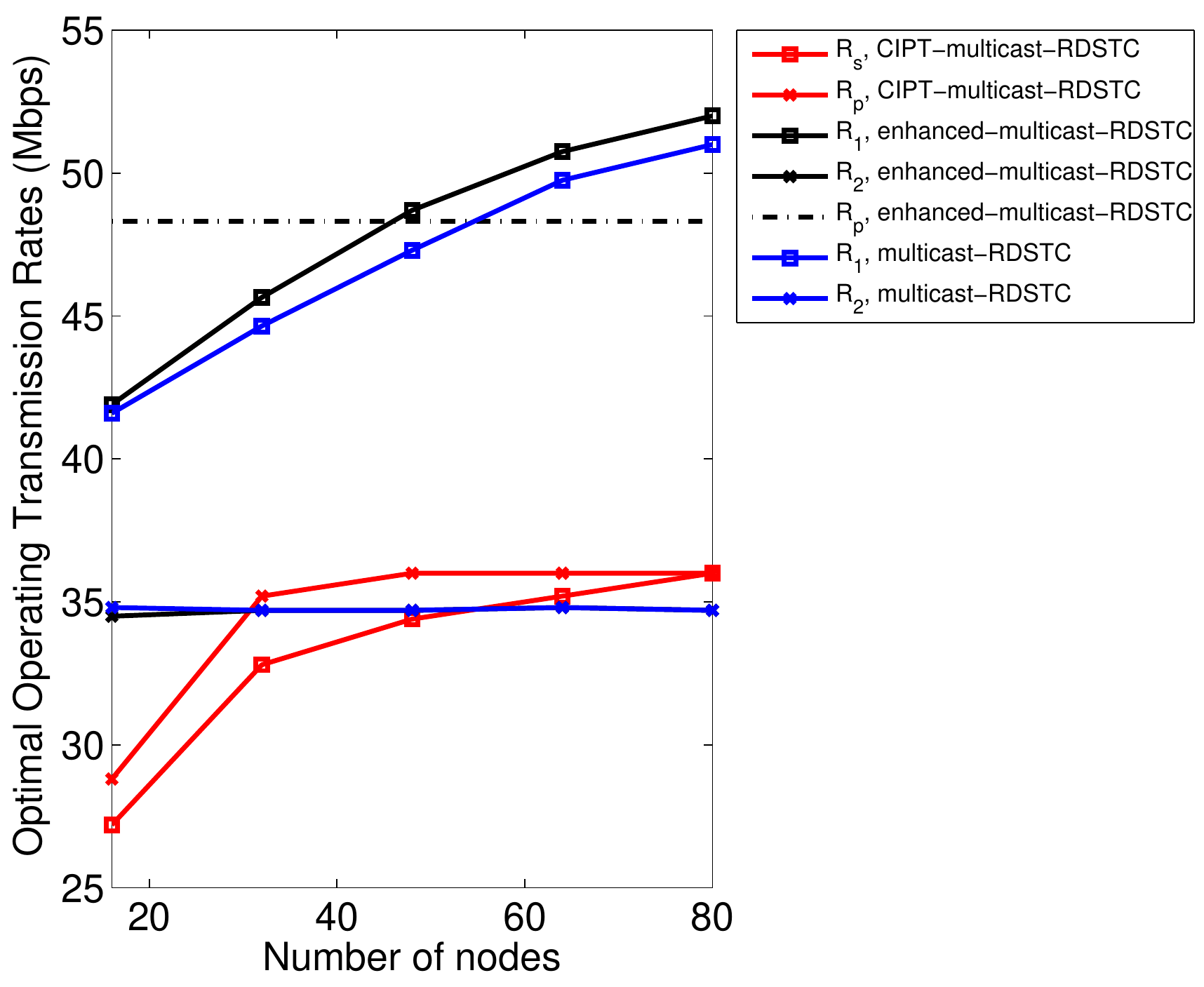}
\else
\includegraphics[width=0.65\columnwidth]{rates_3_movid}
 \fi
\caption{Averaged optimal transmission rates vs number of nodes for different transmission schemes. Number of source packets per FEC block $k=64$.}
\label{fig:rates_opt}
\end{figure}

Fig. \ref{fig:rates_opt} shows optimal transmission rates for the three multicast systems that use R-DSTC. The systems considered in this figure are all with full channel information. \textit{CIPT-multicast} and \textit{enhanced-RDSTC} are using feedback. We can see that the transmission rates for first hop source packets $R_1$ by \textit{multicast-RDSTC} and \textit{enhanced-RDSTC} are substantially higher than $R_s$, the transmission rate for source packets by \textit{CIPT-multicast}. Since the former two systems use relays to help with source packet transmission, the transmission rate of first hop can be higher. Part of the gain of \textit{enhanced-RDSTC} over \textit{multicast-RDSTC} is due to a much higher transmission rate for parity packets $R_p$ than that used for the second hop transmission for the source packet $R_2$.  We believe this is because, typically there are more nodes available for simultaneously sending the parity packets. For the proposed \textit{CIPT-multicast} system, since source packets and parity packets are transmitted only once, to make sure sufficient number of users receives these, transmission rates for both source packets $R_s$ and parity packets $R_p$ are significantly lower. However, even though $R_s$ and $R_p$ are lower, since both source and parity packets are transmitted only once, and the total number of parity packets needed per FEC block $m$ is generally less than the number of source packets, the overall achievable video rate is still significantly higher than the RDSTC-multicast and enhaced-RDSTC-multicast, as illustrated in Fig.\ref{fig_sim}.

\begin{figure}[h]
\centering
\ifCLASSOPTIONtwocolumn
\includegraphics[width=0.95\columnwidth]{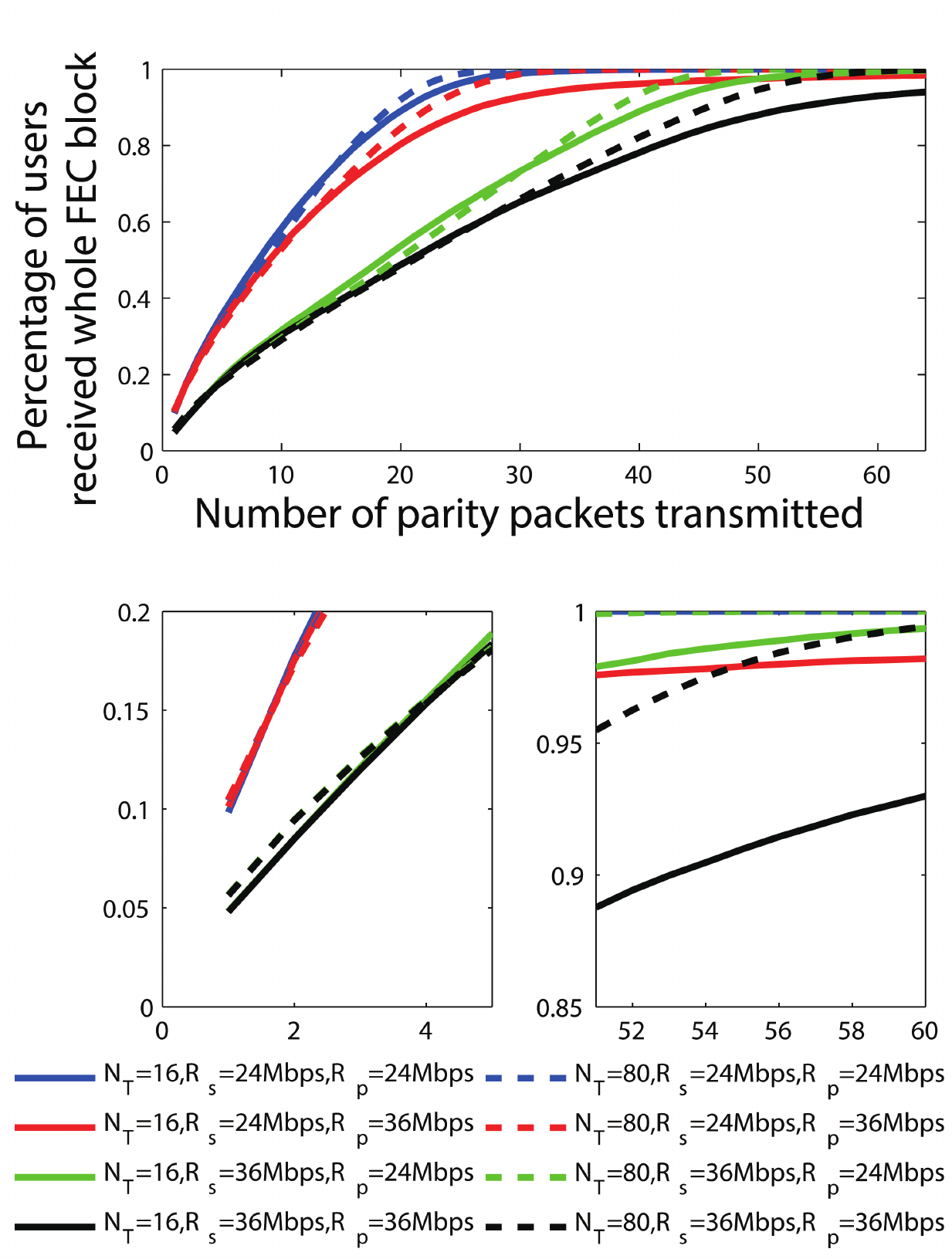}
\else
\includegraphics[width=0.65\columnwidth]{percentage_KM_ex_1115}
 \fi
\caption{Percentage of users that can fully decode a FEC block as the number of parity packets increases under different node counts and different transmission rates. Same color indicates same transmission rates $R_s$ and $R_p$. Solid lines represent $N_T=16$. Dashed lines represent $N_T=80$. The total number of source packets $k$ per FEC block is 64 in this figure.}
\label{fig_relay_trend_whole}
\end{figure}

Fig. \ref{fig_relay_trend_whole} illustrates how does the percentage of users who can decode the entire FEC block increases as more parity packets are transmitted, and the influence of transmission rates and node counts on this percentage. As the upper figure shows, when the transmission rate reduces, the percentage increases faster. This is as expected, as lower transmission rates enable more users to receive $k$ packets earlier.  The figure further shows that, when the transmission rates ($R_s$ and $R_p$) are the same, the curves corresponding to different node counts (curves with the same color) cluster together. This is an advantage for the practical implementation of the proposed system. Recall that the system needs to know the number of relays participating in simultaneous transmission using R-DSTC, to normalize the transmission power. We could predetermine, through simulations, such a curve for each optimal transmission rate combination, which is the same regardless of the number of nodes, and use these pre-determined curves for power normalization during real time transmissions. It is also possible to determine simple mathematical models for such curves, whose parameters are only dependent on the transmission rates, to further simplify the system implementation. 

The lower-left small figure shows that the source packet transmission rate $R_s$ affects the initial number of user who receive the entire FEC block after the source completes its source packet transmission. As the figure shows, at the very beginning, green and black curves (which have the same $R_s$) are overlapping with each other, and so are red and blue curves (which have the same $R_s$). The lower-right small figure shows the impact of parity packets transmission rate $R_p$ on the number of parity packets needed for all users to receive at least $k$ packets. Basically, a larger $R_p$ requires more parity packets. With the same transmission rates combination ($R_s$, $R_p$), when the node count is larger, the system converges faster (i.e. requires fewer parity packets).

\begin{figure}[h]
\centering
\ifCLASSOPTIONtwocolumn
\includegraphics[width=0.95\columnwidth]{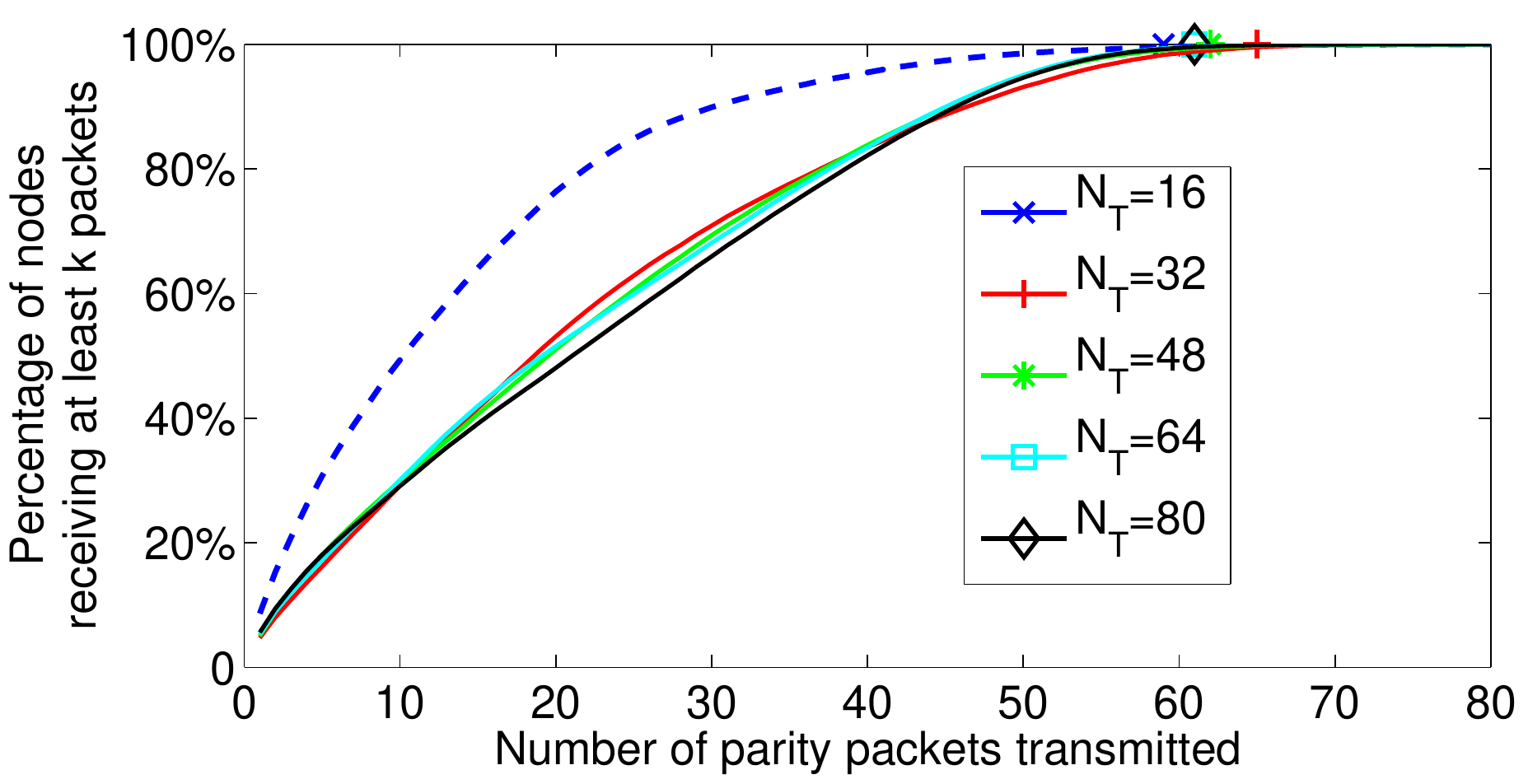}
\else
\includegraphics[width=0.65\columnwidth]{percentage_relay_movid_1115}
 \fi
\caption{Percentage of users that be able to decode whole FEC block vs. number of parity packets transmitted under optimal transmission rates for $k=64$.}
\label{fig_relay_trend_single}
\end{figure}

Figure \ref{fig_relay_trend_single} shows the same type of curves for different node counts, for the case when the system uses the optimal transmission rates for the corresponding node counts. Because the optimal transmission rates for node counts $N_T\ge32$ are similar (cf. Fig. \ref{fig:rates_opt}), these curves are close to each other. The system for $N_T=16$ converges faster (i.e. requires fewer parity packets) because it uses lower transmission rates. However, because it uses substantially lower transmission rates, the total achievable video rate is lower than systems with larger node counts.


\begin{figure}[h]
\centering
\ifCLASSOPTIONtwocolumn
\includegraphics[width=0.95\columnwidth]{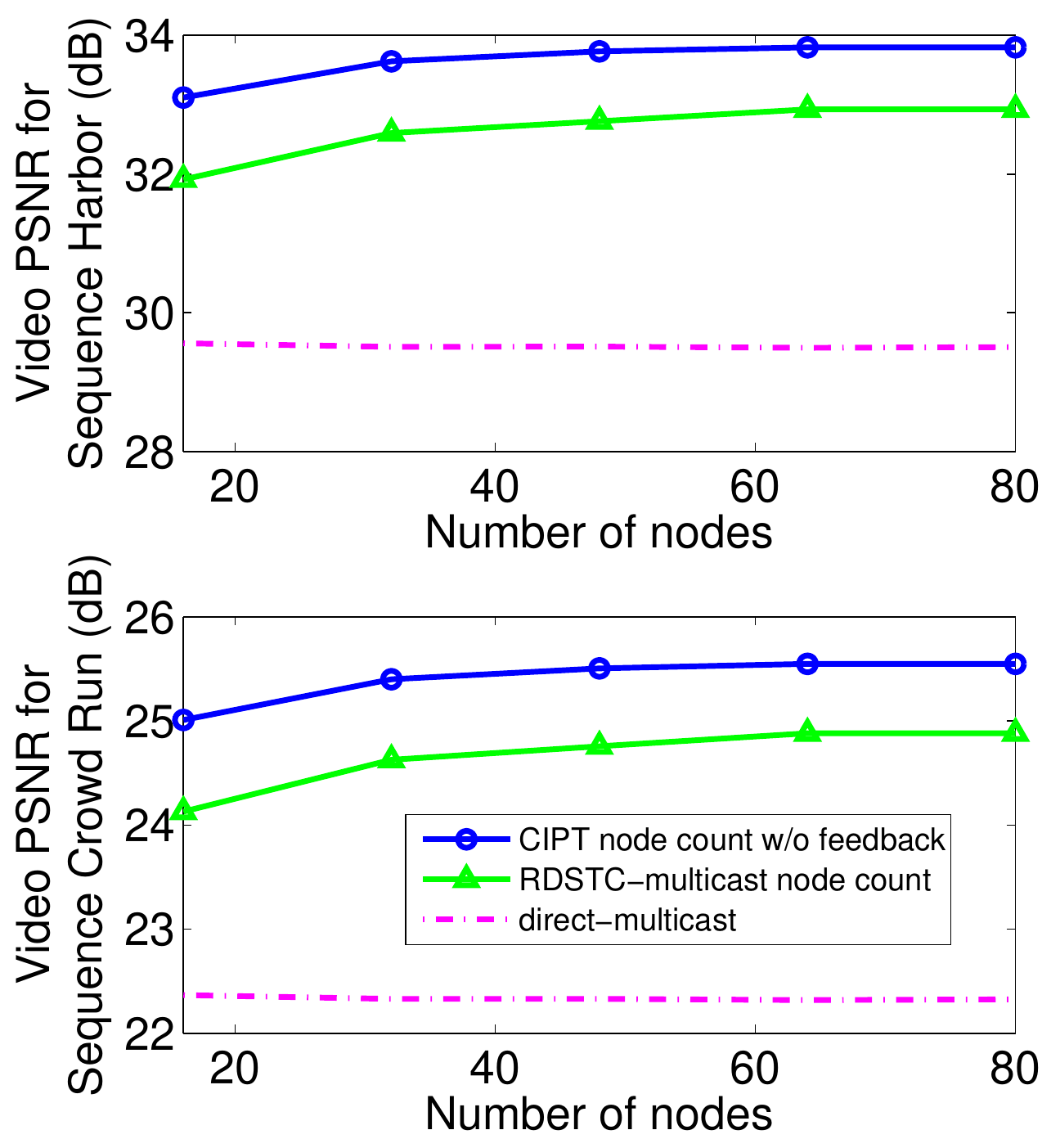}
\else
\includegraphics[width=0.65\columnwidth]{psnr_2_scale}
 \fi
\caption{PSNR of video sequence Harbor (4CIF, 30 Fps) and Crowd Run (1080p, 50 Fps) coded using H.264/AVC at the maximum achievable video rate for different practical multicast systems.}
\label{fig:psnr_nodecount}
\end{figure}

We also want to see performance improvement in terms of video quality over previous systems. Fig. \ref{fig:psnr_nodecount} shows the corresponding PSNR of video sequences Harbour and Crowd Run coded by a H.264/AVC compliant encoder x264\cite{x264} at the maximum achievable video rates by different transmission schemes. Three practical systems are considered in this figure. For \textit{CIPT-multicast} system, we assume there is only node count information available and no feedback exists. Similarly, for \textit{multicast-RDSTC} system, we also assume there is only node count information available. For the sequence Harbour, \textit{CIPT-multicast} has about 1dB gain over \textit{multicast-RDSTC} when $N_T$ is large. When $N_T$ is small, the gain is greater, up to 1.36 dB. For the HD sequence Crowd Run, the gain is between 0.67 to 0.9 db. We would like to note that generally 1dB gain in PSNR is quite noticeable in perceptual quality.

\begin{figure}[h]
\centering
\ifCLASSOPTIONtwocolumn
\includegraphics[width=0.95\columnwidth]{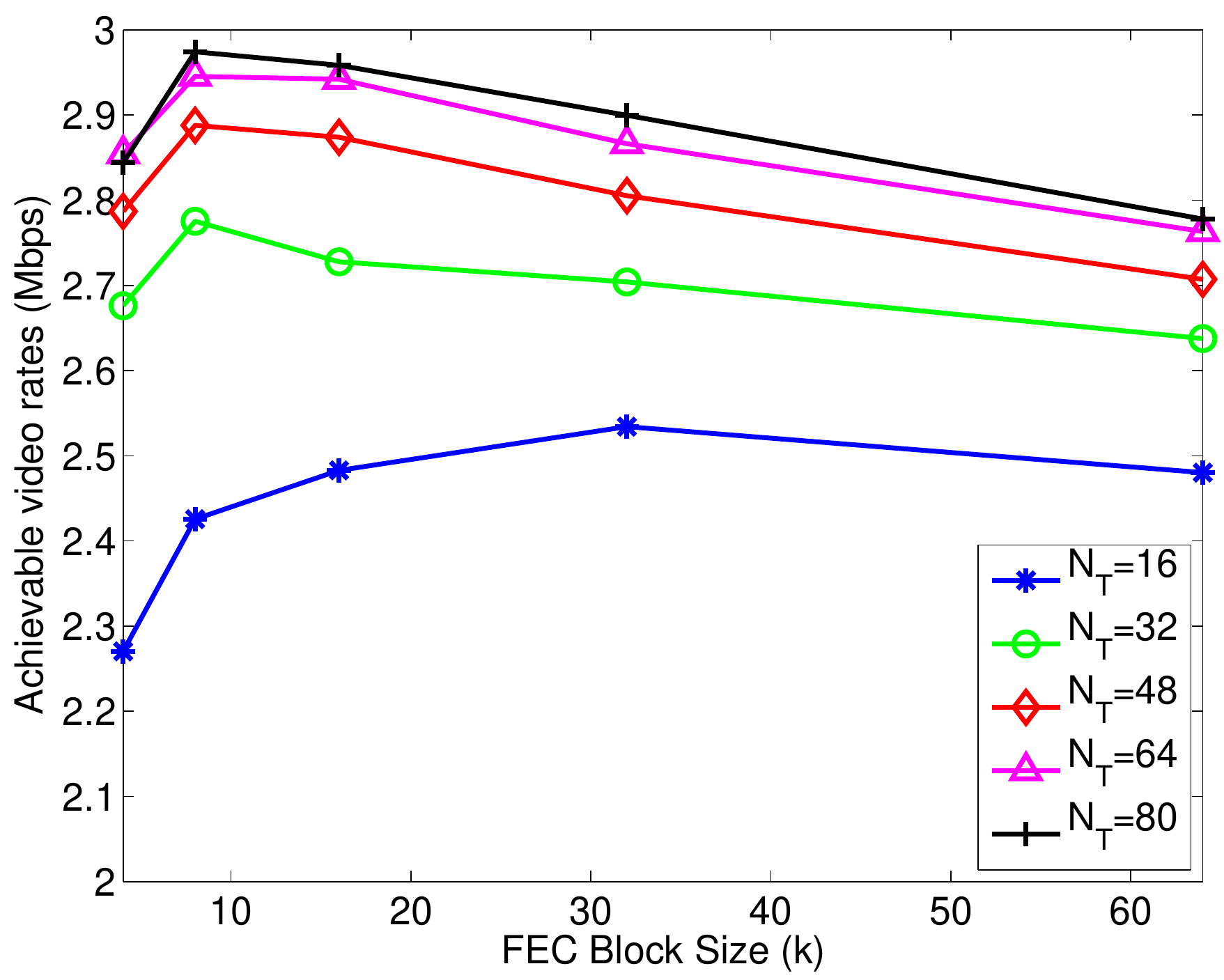}
\else
\includegraphics[width=0.65\columnwidth]{rate_KM_1120}
 \fi
\caption{Achievable video rates for CIPT-multicast full channel with feedback under different FEC block sizes $k$ and node counts $N_T$.}
\label{fig:rates_K}
\end{figure}

\begin{figure}[h]
\centering
\ifCLASSOPTIONtwocolumn
\includegraphics[width=0.95\columnwidth]{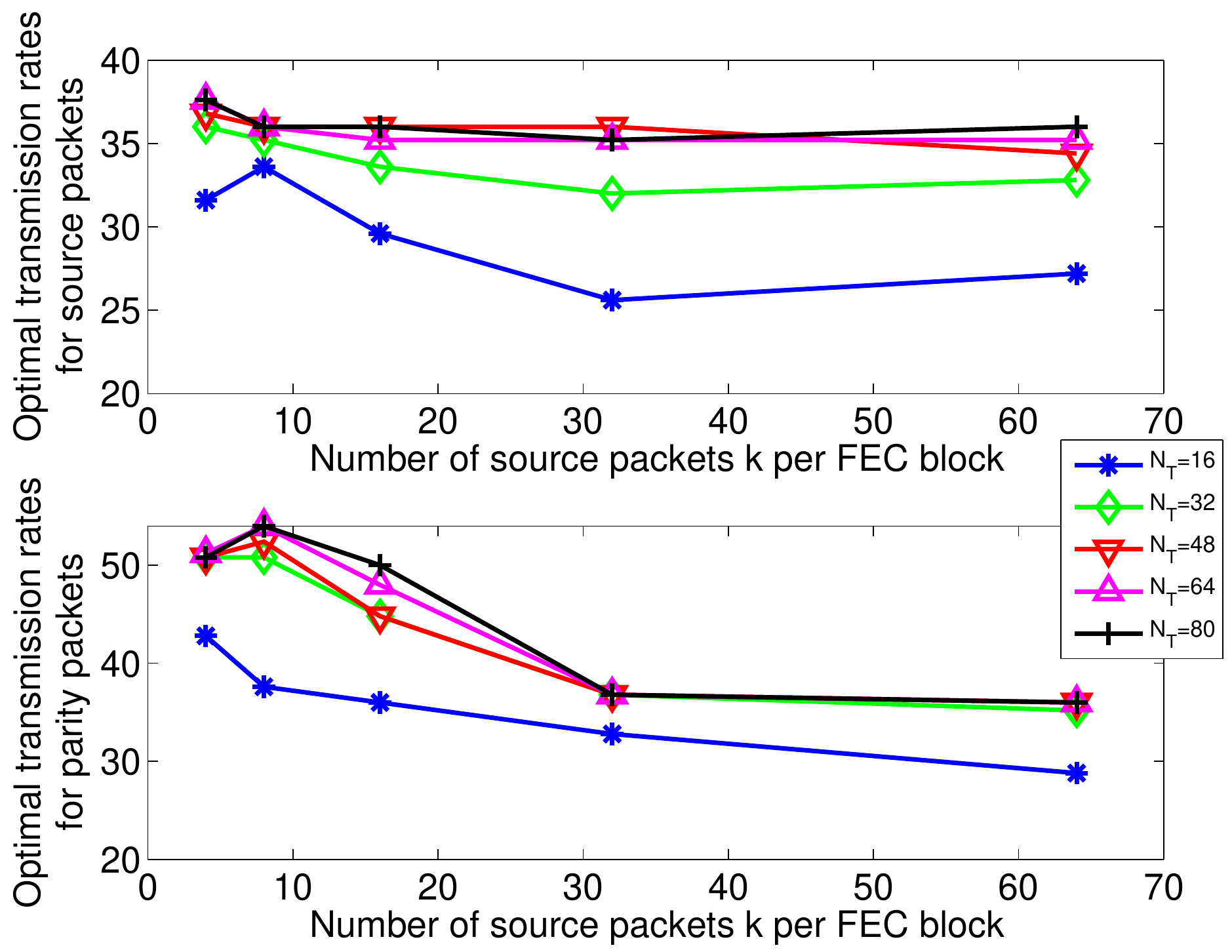}
\else
\includegraphics[width=0.65\columnwidth]{rates_opt_KM_1115}
 \fi
\caption{Optimal transmission rates for CIPT-multicast full channel with feedback for different node counts with different FEC block size.}
\label{fig:rates_K_opt}
\end{figure}

\begin{figure}[h]
\centering
\ifCLASSOPTIONtwocolumn
\includegraphics[width=0.95\columnwidth]{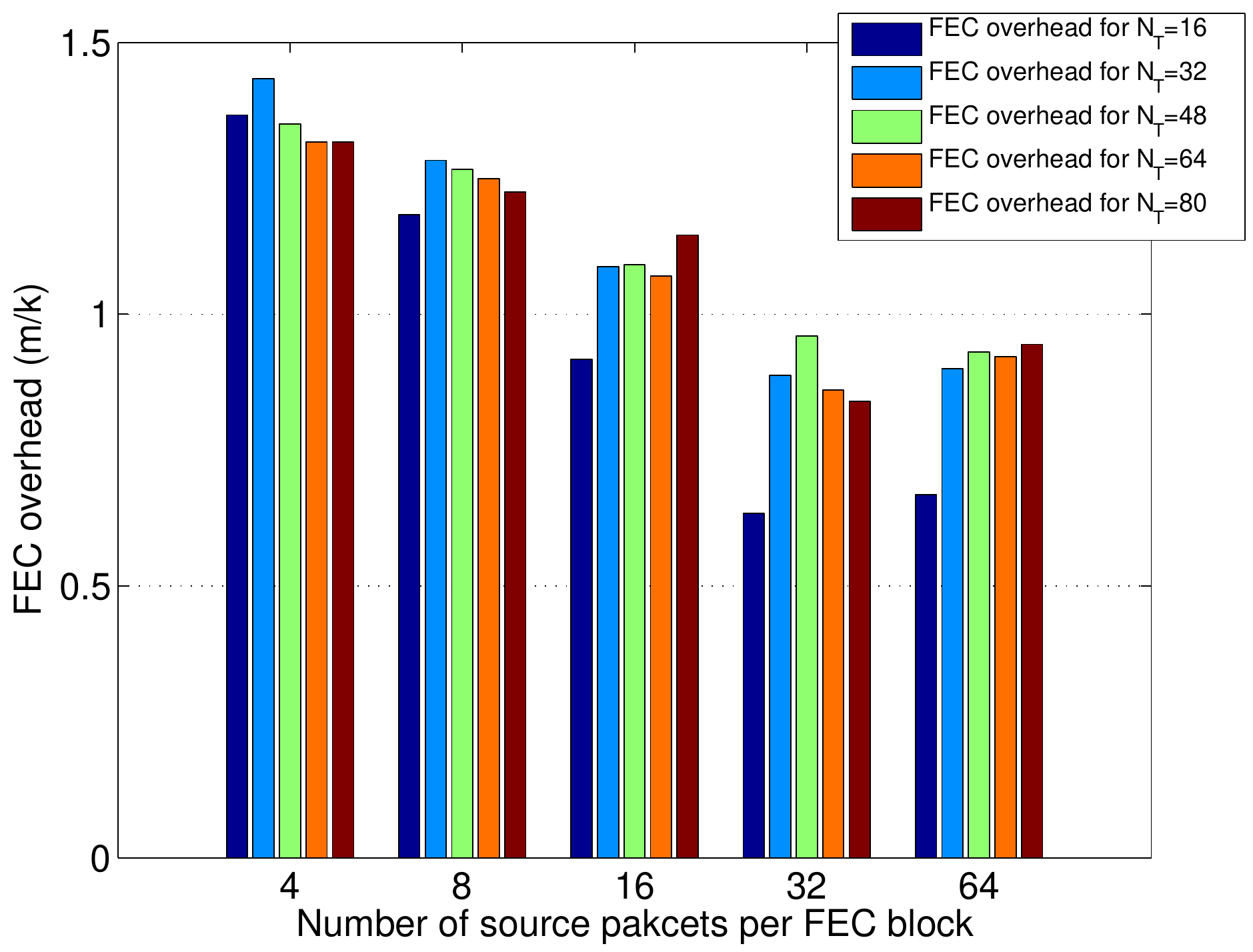}
\else
\includegraphics[width=0.65\columnwidth]{bar_KM_1120}
 \fi
\caption{FEC overhead $m/k$ needed for different FEC block size and different node count. Same color indicates the same node count. Each group of bars have the same FEC block size.} 
\label{fig_bar_km}
\end{figure}

Previous discussions focused on the case when the number of source packets in a FEC block is fixed at $k=64$. Next, we examine the impact of $k$ on the system performance. Fig. \ref{fig:rates_K} shows the optimal achievable video rate for different $k$ and node count $N_T$. For the same $k$, the system performance in terms of $R_v$ still follows the same trend of Fig. \ref{fig_sim}, with increasing performance for larger $N_T$. For the same $N_T$, the best performance occurs at some intermediate $k$, depending on $N_T$. This may be contradictory to the conventional wisdom that FEC is more effective with a larger block length. However, this is because the block size $k$ affects two different aspects of the CIPT system. On the one hand, when $k$ gets bigger, at the end of source packets transmission, fewer user will likely receive all $k$ source packets. As a result, fewer users will be able to participate in the parity packet transmission, necessitating lower parity transmission rates. In order to increase the number of initial relays, a system using a larger $k$ also needs to lower the source transmission rate, especially for systems with lower node count.  This is demonstrated in Fig. \ref{fig:rates_K_opt}, which shows that the optimal source and parity transmission rates both decrease with $k$. On the other hand, with a larger $k$, the FEC is more effective in that it requires fewer parity overhead (defined by the ratio $m/k$). This is demonstrated in Fig. \ref{fig_bar_km}, which shows that a higher overhead is needed for a system with smaller $k$. Because the overall video rate increases with transmission rates but decreases with the overhead rates, some intermediate $k$ is optimal. Figure \ref{fig_bar_km} further demonstrates that under the optimal transmission rates, the necessary overhead rate, $m/k$, is smaller than 1.5 for different $k$ and different $N_T$. For $k > 32$, the ratio $m/k$ is below 1.

%% file: cipt_con.tex
\section{Conclusion}
\label{sec:con}
In this paper, we propose an innovative wireless video multicast system which features cooperative incremental parity packet transmission using R-DSTC. Both source packets and parity packets are transmitted using only one hop. Users who receive $k$ packets will generate parity packets and join parity packet transmission using R-DSTC. In the most ideal but also most complex implementation, the system periodically updates the full channel information through message exchange among all nodes and uses feedback from users to determine whether to continue sending parity packets. We optimize the transmission rates for both the source packets and parity packets for each possible channel state corresponding to the full channel information. Three suboptimal but more practical schemes are also investigated. One does not require the full channel information but requires feedback, one requires the full channel information but not the feedback, and the third one requires neither full channel information nor the feedback. We show that all different variations of the proposed CIPT scheme provides substantially higher video rates over the \textit{multicast-RDSTC} scheme in which both source and parity packets go through a two-hop transmission. The CIPT system also provides significant gain over the enhanced-multicast-RDSTC system, which requires feedback. Our simulation results further show that the feedback information is more important than the full channel information in maintaining a high achievable rate.




There are multiple possible paths for future research. One is to further improve the performance of this proposed scheme by discontinuing relays from parity packet transmissions after they have transmitted a certain number of times. This could be beneficial because these nodes may not be helpful any more, and yet by removing them from parity transmission, the remaining parity relays can transmit at higher power. Another way to improve the performance is by dynamically increasing the parity transmission rate as more parity packets are sent. This is likely to be beneficial because, as more relays join parity transmission, higher transmission rates are sustainable. Another research direction is to adapt the proposed scheme to layered coded video, as done in \cite{OZGU:tmm11}. This will deliver differentiated quality to different nodes. For example, we can choose the parity packet number for the base layer video so that all users can recover all  source packets with a high probability, and choose the parity packet number for the enhancement layer so that only a certain percentage of users can recover all source packets.